 \documentclass[aps,prl,twocolumn,showpacs]{revtex4}
\usepackage{graphicx}
\usepackage{amssymb}
\usepackage{amsmath}
\begin{document}

\title{Comment on ``Self-organized criticality and absorbing states:
Lessons from the Ising model"}

\author{Mikko J. Alava$^1$, Lasse Laurson$^1$,
Alessandro Vespignani$^2$ and Stefano Zapperi$^3$}

\affiliation{$^1$ Laboratory of Physics,Helsinki University
of Technology,
FIN-02015 HUT, Finland}
\affiliation{$^2$ School of
Informatics and Department of Physics,
Indiana University, Bloomington, IN 47408, USA}
\affiliation{$^3$ INFM-CNR, S3, Dipartimento di Fisica, Universit\`a
di Modena e Reggio Emilia, via Campi 213/A, I-41100, Modena, Italy
and ISI Foundation, Viale S. Severo 65, 10133 Torino, Italy }

\begin{abstract}
According to Pruessner and Peters
[Phys. Rev. E {\bf 73}, 025106(R) (2006)], the
finite size scaling exponents of the order
parameter in sandpile models depend on the tuning
of driving and dissipation rates with system size.
We point out that the same is not true for
{\em avalanches} in the slow driving limit.
\end{abstract}

\maketitle

\date{\today}

In a recent paper Pruessner and Peters investigated the relation
between self-organized criticality (SOC) in sandpile models and
absorbing state phase transitions, on the basis of an analogy with
standard equilibrium critical phenomena, in particular the
two-dimensional Ising model \cite{pruesner2006}. According to Ref.
~\cite{pruesner2006} only a careful choice of the system size
dependence of the driving and dissipation parameters would yield the
scaling results of the underlying phase transition. Here we point
out that this reasoning does not apply in particular when one
studies ``SOC variables'' such as various measures of avalanches in
the slow driving limit, as is traditionally done in this context.
This is confirmed by numerical simulations presented below.

In SOC sandpile models, the steady-state is maintained
by a balance of dissipation $\epsilon$ and driving $h$.
In particular, the control parameter of the absorbing
phase transition, i.e. the average height of the pile
$\zeta$, evolves on the average as
\begin{equation}
\dot{\zeta}= h -\epsilon \rho,
\label{eq:soc}
\end{equation}
where $\rho$ is the density of active sites, the order parameter of
the absorbing phase transition. In the limit $h\to 0^+$,$\epsilon
\to 0^+$, the control parameter flows to the critical value
$\zeta_c$ and the model shows scale invariance \cite{dickman2000}.
The question raised by the authors of Ref.~\cite{pruesner2006} is
how to apply finite-size scaling (FSS) to the order parameter $\rho$
if $\epsilon (L) \sim L^{-\kappa}$ and $h (L) \sim L^{-\omega}$ are
taken to be functions of the system size $L$. To investigate this
issue, Pruessner and Peters employ a similar ``driving'' to the
Ising model, using a fluctuating inverse temperature $\theta$
evolving as
\begin{equation}
\dot{\theta}= h -\epsilon |m|,
\label{eq:ising}
\end{equation}
where $|m|$ is the absolute value of the magnetization. Using this
driving, the system size dependence of the order parameter, $\langle
|m| \rangle \sim L^ {\kappa-\omega}$, coincides with that expected
from standard FSS only when $\omega-\kappa=\beta/\nu$, where $\beta$
and $\nu$ are the equilibrium exponents for the order parameter and
the correlation length, respectively. We notice that the case
$\omega-\kappa>\beta/\nu$ corresponds precisely to the slow driving
limit of SOC, where there is complete time-scale separation between
driving and dissipation (i.e. for most purposes one may even take
$\omega \to \infty$ and wait an infinite amount of time after each
driving event). In this limit, the effective temperature defined in
Ref.~\cite{pruesner2006} also would diverge. The authors, in analogy
with the Ising model, conclude that the slow driving SOC state
should {\it not} correspond to the critical point of an absorbing
phase transition, in contradiction with the evidence from numerical
simulations of sandpile models \cite{lubeck2004}.

The apparent contradiction disappears when we notice that
avalanche statistics, as typically studied in the case of
SOC, and the average order parameter studied in
Ref.~\cite{pruesner2006} are not equivalent measures of the
criticality of the underlying absorbing state phase transition.
As pointed out by Pruessner and Peters, it is indeed possible
to tune the system size dependence of the order parameter,
or average activity, by choosing the size scaling of
driving and dissipation rates appropriately. In the slow
driving limit relevant for SOC, however, this is just a
trivial consequence of the drive rate dependence of the
quiescent periods between avalanches. The avalanches
themselves are not affected by the drive rate
in any way as long as it is slow enough such that no new
grains are added while the system is active.

In SOC sandpiles, one usually implements open boundary conditions
and infinitely slow drive, corresponding to $\kappa =2$ and $\omega
\rightarrow \infty$. The mappings to absorbing phase transitions and
to depinning transitions allows to obtain the scaling behavior for
any value of $\kappa$, provided that we remain in the time scale
separation regime \cite{dickman2000}. In general, sandpile models
exhibit FSS forms for the avalanche sizes $s$, of the type
\begin{equation}
P(s,\xi) = s^{-\tau_s} P(s/\xi^D_s) .
\label{eqn:fss}
\end{equation}
$\tau_s$ and $D$ are critical exponents related to the
underlying depinning transition and $\xi(\omega,\kappa)$ is
the cut-off scale that is determined by the condition of
balance between dissipation and drive (which also results
in the steady state condition $\rho = h/\epsilon$). One
transparent argument to compute $\xi$ is to look at the
dynamics following the addition of a single grain, which
gives rise to an avalanche of average size $\langle s\rangle$
dissipating on the average one grain \cite{tang88}. Thus
one has the condition $\epsilon(L) \langle s \rangle (\xi) = 1$
and obtains
\begin{equation}
\xi \sim L^{\frac{\kappa}{D(2-\tau_s)}}
\label{eqn:xi}
\end{equation}
Notice in particular that $\xi$ is not dependent on the
drive-rate or $\omega$.

To confirm this, we consider the Manna sandpile model
with periodic boundary conditions, slow driving (i.e.
$\omega=\infty$) and bulk dissipation. The model is studied
here on a 2d lattice, where for each lattice site $i$ one
assigns an integer variable $z_i$ (the number of ``grains'').
If $z_i > z_c = 1$, a ``toppling'' occurs and the grains
are redistributed according to $z_i \rightarrow z_i-2$
and $z_{nn}=z_{nn}+1$, where $z_{nn}$ are two randomly
chosen nearest neighbors of site $i$. The dissipation is
implemented by removing a toppling grain from the system
with probability $\epsilon \propto L^{-\kappa}$. Here we
consider the two cases with $\kappa=1$ and $\kappa=3$,
respectively. In Figure 1 we show that the scaling follows
the predictions of absorbing phase transitions: the cutoff
scale of the avalanche size distribution scales according
to Eq. (\ref{eqn:xi}), regardless of the $L$-dependence of
$\epsilon$.

To summarize, we would like to point out that the
conclusions of Pruessner and Peters are misleading in the
sense that avalanche statistics in sandpile models
indeed follows from the underlying absorbing state
transition whenever one studies the slow driving limit
 $\omega-\kappa > \beta/\nu$. This is the condition of
complete time scale separation between driving and
avalanche propagation that has been recognized already in
the early literature as the crucial ingredient for SOC
in sandpile models \cite{grinstein1995}.

\begin{figure}
\includegraphics[width=9cm]
{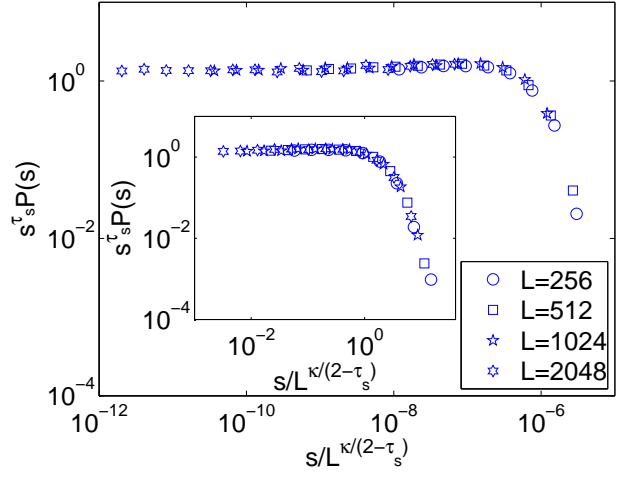}
\caption{Scaling plots of the avalanche size distributions
from the two-dimensional Manna model with bulk dissipation.
The inset presents the collapse in the case $\kappa=1$, while
in the main figure the case $\kappa=3$ is shown. Both collapses
correspond to the value $\tau_s = 1.28$.}
\label{fig1}
\end{figure}


\begin{thebibliography}{10}
\bibitem{pruesner2006}
G. Pruessner and O. Peters
 Phys. Rev. E 73, 025106(R) (2006).
\bibitem{dickman2000}
R. Dickman, M. A. Mu{\~n}oz, A. Vespignani, and S. Zapperi, Braz. J.
Phys. 30, 27 (2000); M. Alava, J. Phys. Cond. Matt.  14, 2353
(2002).
\bibitem{lubeck2004}
S. L\"ubeck, Int. J. Mod. Phys. B {\bf 18}, 3977 (2004).
\bibitem{tang88}
C. Tang and P. Bak, Phys. Rev. Lett. 60, 2347 (1988).
\bibitem{grinstein1995}
G. Grinstein, in Scale Invariance, Interfaces and
Non-Equilibrium Dynamics, Vol. 344 of NATO Advanced
Study Institute, Series B: Physics, edited by A. McKane
et al. (Plenum, New York, 1995).
\end{thebibliography}
\end{document}